\documentclass[preprint,showpacs,preprintnumbers,amsmath,amssymb]{revtex4}
\usepackage{graphicx}
\usepackage{dcolumn}
\usepackage{bm}
\begin{document}
\title{Structure and organization in inclusion-containing   bilayer membranes}

\author{Chun-Lai Ren and Yu-qiang Ma}
\altaffiliation[ ]{Corresponding author: myqiang@nju.edu.cn}
\address{National Laboratory of Solid State Microstructures, Nanjing University, Nanjing 210093, China}

\begin{abstract} Based on a considerable amount  of experimental evidence for generic
  properties of lateral organization of lipid membranes containing
  inclusions, we first present
  a general model system of   bilayer membranes embedded by  nanosized inclusions,
  and  account well for a series of unexpected behaviors in related
experimental findings. (1) The appearance and disappearance of
lipid/inclusion-rich rafts are observed with increasing  the
  inclusion content. (2) The chain arrays of inclusions
may form at high    concentrations. (3) Location of inclusions
changes with increasing the   inclusion content, and may undergo a
layering transition from one-layer located in the center of the
bilayer to two-layer structure arranged in opposing leaflets of a
bilayer.  (4) The membrane fluidity  is enhanced by the presence
of inclusions. Our theoretical predictions address the complex
interactions between membranes and inclusions, suggesting a
unifying mechanism which reflects the competition between the
conformational entropy of lipids
 favoring  the formation of lipid-rich rafts and the
steric repulsion of inclusions leading to the uniform dispersion.
The present study advances our understanding of membrane
organization by unifying these experimental evidences of   real
biomembranes with inclusions  which can be different, but with the
hydrophobic and rigid properties.
\end{abstract}

 \maketitle
Recently, there is growing evidence that due to the presence of
inclusions within membranes, the distribution of lipids is
inhomogeneous, where lateral segregation could induce the
formation of lipid/inclusion-rich raft domains.   For instance,
cholesterol which is one of the most important regulators of lipid
organization, prefers to have conformationally ordered lipid
chains next to it due to its hydrophobically smooth and stiff
steroid ring structure\cite{ourit1,ne20}, and promotes the
formation of lipid rafts(see, for example,
reviews\cite{ourit1,ourit2,renew4} and recent research
works\cite{mmm4}). On the other hand, it was reported \cite{ne9}
that hydrophobic drugs such as taxol(paclitaxel)\cite{ne8,ne9} and
dipyridamole(DIP)\cite{ne6}) may assist the formation of
lipid/drug-enriched raft domains, and increase with increasing the
drug content, but disappear at high concentrations, which has also
been observed in cholesterol-lipid systems \cite{renew5}.
Particularly, the perturbed lipids may lead to chaining of
cholesterols \cite{renew5,roger} or drugs \cite{ne5,ne9} inside
the bilayer. Furthermore,    the introducing of taxol drug into
the lipid layer will perturb the hydrocarbon chain conformation,
which may increase membrane fluidity\cite{ne5,ne8,ne9}. Most
recently, some foreign inclusions such as silver nanoparticles
were reported to have similar effects on the membrane
fluidity\cite{a4}.

Despite the common properties of membrane organization due to
distinct inclusions and despite its important significance in
cellular functions such as signal transduction  and membrane
trafficking\cite{ourit2},  the influence of the inclusions on such
a change in lateral organization has not yet been considered in
computational and theoretical investigations, and further insight
into the mechanisms behind general evidence from lipid-inclusion
complexes remains poor\cite{schick5}. Previous theoretical works
were concerned with the hydrophobic mismatch interaction between
inclusions\cite{bsmit,brr,bsd} and the possible formation of lipid
rafts\cite{ourit1,schick5}. However, a detailed structural change
with varying the inclusion content has not been systematically
elucidated. In this letter, we examine a simple model that not
only allows us to present a unifying description of these
phenomena with varying  inclusion concentrations, but also sheds
light on physical mechanism behind membrane organization due to
distinct inclusions such as intrinsic membrane protein,
cholesterol, hydrophobic drug, or other bionanoparticles.

Consider a lipid bilayer membrane containing   $n_d$  inclusions
of radius R in an aqueous environment(Fig. 1).  The volume of
system is $V=L_x\times L_y\times L_z$, where $L_x$ and $L_y$ are
lateral membrane lengths under
 periodical boundary conditions, and $L_z$ is the size of system
 along the membrane normal direction. The bilayer membrane is composed of
one type of lipids with two hydrophobic tails.  The number of
lipids
 is given by $n_{l}=2\times \sigma \times
L_x\times L_y$, where $\sigma$ is the number density of lipids in
one leaflet. The headgroup of lipids has the volume $v_{h}$, and
two equal-length tails composed of $ N/2 $ segments each are
assumed as flexible Gaussian chains\cite{gh}. The segment has the
volume $\rho_{0}^{-1}$ and the length $a$. Thanks to the smooth
and rigid property of bioinclusions compared with the flexible
lipids\cite{ourit1,renew4,bsmit}, we assume inclusions as
\begin{figure}
\setlength{\abovecaptionskip}{0.5pt}
\includegraphics[width=5cm]{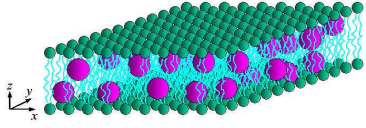}
\caption{(color) A schematic   of the lipid bilayer membrane
containing inclusions.}
\end{figure}
hard-sphere nanoparticles. The concentrations of the head and
tails of lipids are $\phi_{h}=n_{l}v_{h}/V$ and
$\phi_{t}=n_{l}N\rho _0^{-1}/V$, and the inclusions $\phi_{d}=
4\pi R^{3}n_d/3V$. The  solvent molecules have the volume $v_{s}$
and the concentration $\phi_{s}=1-\phi_{t}-\phi_{h}-\phi_{d}$
\cite{a5}. Recently, the self-consistent field theory (SCFT) has
been proven to be powerful in calculating equilibrium morphologies
in polymeric systems\cite{add3,m3,em3}, while nanoparticles can be
treated by density-functional theory (DFT) \cite{add1} to account
for steric packing effects of particles. Interestingly, Thompson
et al. \cite{add3} developed a hybrid SCFT/DFT approach to study
mixtures of diblock copolymer and nanoparticles.  On the other
hand, the SCFT method is extended to  successfully study the phase
behavior of pure lipid systems\cite{gh,leer3, schick5}.
 Here, we will extend the hybrid SCFT/DFT approach\cite{add3} to
calculate the structural organization of the   bilayer membrane in
the presence of inclusions. The resulting free energy $F$ for the
present system \cite{dfg} is given by
\begin{eqnarray}
\frac{NF}{\rho _0 k_B T V} &=&-\phi_{t}\ln (\frac{Q_{l}}{V\phi_{t}})-\frac{\phi_{s}}{\alpha_{s}}\ln (\frac{Q_{s}}{V%
\phi_{s}})-\frac{\phi_{d}}{\alpha}\ln (\frac{Q_{d}\alpha}{V\phi_{d}})   \nonumber\\
 &&+\frac 1V\int
d\mathbf{r}[\chi _{th}N\varphi _t({\bf r})\varphi _h({\bf r})%
+\chi _{ts}N\varphi _t({\bf r})\varphi_s({\bf r}) \nonumber\\
&&+\chi _{td}N\varphi _t({\bf r})\varphi_d({\bf r})
+\chi _{hs}N\varphi _h({\bf r})\varphi_s({\bf r})%
  \nonumber\\
&&+\chi _{hd}N\varphi _h({\bf r})\varphi_d({\bf r})+\chi
_{sd}N\varphi _s({\bf r})\varphi_d({\bf r})\nonumber\\
&&-w_t({\bf r})\varphi _t({\bf r})-w_h({\bf r})\varphi _h({\bf r})-w_s({\bf r})\varphi_s({\bf r})%
  \nonumber\\
&&-w_d({\bf r})\rho_d({\bf r})-\xi({\bf r}) (1-\varphi _{t}({\bf
r}) - \varphi _{h}({\bf r})\nonumber\\
&&-\varphi _{s}({\bf r})-\varphi _{d}({\bf r}))+\rho_{d}({\bf
r})\Psi_{hs}(\overline{\varphi }_{d})]\;,
\end{eqnarray}
where  $\alpha_{s}= v_{s}\rho_{0}/N$, and $\alpha=4\pi
R^{3}\rho_{0}/3N$.   $\chi _{th}$, $\chi _{ts}$, $\chi _{td}$,
$\chi _{hs}$, $\chi _{hd}$, and $\chi _{sd}$ are the Flory
interaction parameters between tail-head, tail-solvent,
tail-inclusion, head-solvent, head-inclusion, and
solvent-inclusion, respectively. $\varphi _{t}(\bf r)$, $\varphi
_{h} (\bf r)$, $\varphi _{d}(\bf r)$, and $\varphi _{s}(\bf r)$
are the local volume fractions of lipid tail, head group,
inclusion, and solvent, and $w_{t}(\bf r)$, $w_{h}(\bf r)$,
$w_{d}(\bf r)$, and $w_{s}(\bf r)$ are corresponding
self-consistent fields. $\xi(\bf r) $   ensures the
incompressibility of the system, and $\rho_d(\bf r)$ stands for
the inclusion center distribution.  $Q_l $,
  $Q_{s} $, and $Q_{d} $ are respective partition functions
for lipid, solvent, and inclusions\cite{add3}. The last term in
Eq. (1) is the nonideal steric interaction term \cite{add1} with
the weighted inclusion density $\overline{\varphi }_{d}(\bf r)$
\cite{add3}. The   fields and densities are then determined by
minimizing the free energy  in Eq.(1), and the resulting
self-consistent equations   can be solved numerically\cite{add3}.
To reasonably describe
  the interactions of
hydrophobic inclusions dispersed in bilayer lipids with
hydrophilic heads and hydrophobic tails, we choose  $\chi
_{hs}N=0$, $\chi _{td}N=3.5$, $\chi _{hd}N=15.0$, $\chi
_{ts}N=15.0$, $\chi _{sd}N=20.0$,  and $\chi _{th}N=25.0$. The
other parameters are fixed to be $N=30$, $v_{h}=v_{s}=6\rho
_0^{-1}$, $\sigma=0.08$,
   $L_x=L_y=60a$, and $L_z=15a$. Here, the chosen $L_z$ was  large enough, ensuring the  solvent
concentration $\phi_{s}=1$ at $z=0$ and $L_z$.

\begin{figure}
\setlength{\abovecaptionskip}{0.3pt}
\includegraphics[width=6.5cm]{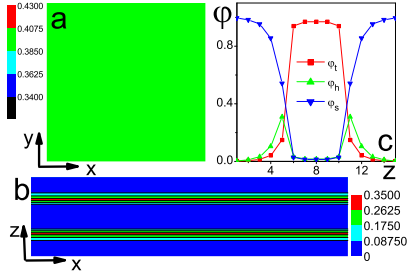}
\caption{(color) The pure bilayer membrane in an aqueous
environment. (a) The top view of averaged density distribution of
lipid tails in the upper leaflet. (b) laterally averaged density
profiles of solvent($\varphi_{s}$), lipid tail($\varphi_{t}$), and
head group($\varphi_{h}$) along z axes. (c) The x-z cross-section
of density distribution of lipid heads averaged in the
y-direction.}
\end{figure}

We first examine the case of a  bilayer  membrane in an aqueous
environment in the absence of inclusions. Figure 2a shows the
lateral distribution of lipid tails in one leaflet, which is
uniform. The distribution is symmetrical with respect to the
opposing  leaflet. Figure 2b provides the average distribution of
headgroups in the x-z cross-sections, reflecting that the shape of
membrane surfaces is smooth in the absence of inclusions, since
the  lipid  length is matched. In this case, the membrane is in a
gel phase for the limited lateral mobility\cite{renew5,ne5}, where
lipid tails tightly pack and hardly move in the pure membrane.
Figure 2c shows the laterally averaged density profiles of
$\varphi_t$, $\varphi_h$, and $\varphi_s$ across a bilayer, which
displays the basic structure of the bilayer and is favorably
comparable with mesoscopic simulation and other coarse-grained
models describing the bilayer composition\cite{leer3,xe1}. This
validates the used SCFT which can reasonably explore
conformational properties of lipids, in contrast to
phenomenological models that ignore much of the internal structure
of the bilayer\cite{schick5}.  Particularly, the  approach will
become  powerful in exploring the lateral inhomogeneity of
membrane with inclusions, in contrast to other field theory or
simulation schemes\cite{schick5},  where the only one-dimensional
density profiles   along the membrane normal is shown in the cost
of including molecular details of lipids.
\begin{figure}
\setlength{\abovecaptionskip}{0.3pt}
\includegraphics[width=8cm]{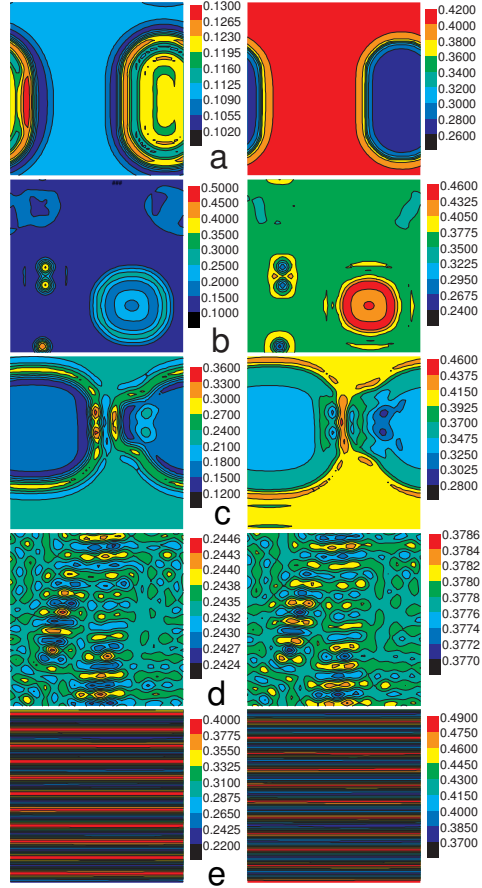}
\caption{ (color)  The top views of averaged density distributions
of inclusions (left) and   lipid tails (right) in upper leaflet.
(a) $\phi_{d}=0.10,$ (b) $\phi_{d}=0.15,$ (c) $\phi_{d}=0.20,$ (d)
$\phi_{d}=0.24,$ and (e) $\phi_{d}=0.32$. Here we fix  $R=2.5a$
which is smaller than the membrane thickness.}
\end{figure}
\begin{figure}
\setlength{\abovecaptionskip}{0.3pt}
\includegraphics[width=9cm]{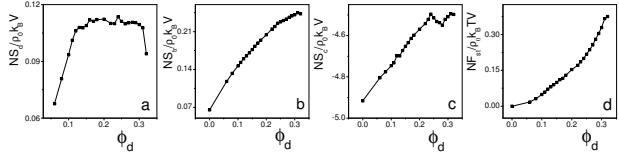}
\caption{ (a) The translational entropy $S_{d}$ of inclusions, (b)
The translational entropy $S_{tr}$ of lipids, (c) the
conformational entropy $S_c$ of lipids, and (d) the steric energy
$F_{st}$ of inclusions vs the inclusion concentration. }
\end{figure}

Figure 3 shows the  in-plane density distributions of inclusions
in the left column and the lipid tails in the right column with
increasing the inclusion concentration $\phi_d$. When $\phi_d$ is
low, the translational entropy of inclusions has a significant
contribution to the free energy of the system. Any compositional
fluctuation of inclusions may lead to the lateral inhomogeneity of
lipid composition, in which lipids are depleted in the
inclusion-rich regions for ensuring the large translational
entropy of inclusions(Fig. 3a). Astonishingly, as  $\phi_d$ is
increased, the lipid/inclusion-rich rafts appear(Fig. 3b). In this
case, the entropic contribution of lipids becomes significant, and
tail chains closing to the rigid surfaces of inclusions can get
extra conformational flexibility\cite{ourit1,brr,renew005,a4},
which enriches the lipids surrounding inclusions. Therefore,
inclusions are accumulated in certain membrane regions, which
leads to the formation of lipid-rich rafts. The  raft size  may be
enlarged by increasing $\phi_{d}$ (Fig. 3c). Interestingly, as
$\phi_{d}$ is added to a certain range($\phi_{d}= 0.24 \sim0.28$),
lipid-rich rafts disappear,  but instead the uniform distribution
of both inclusion and lipid tail occurs (Fig. 3d). This unexpected
behavior provides a strong support for the experimental findings
in drug-membrane \cite{ne9} and cholesterol-membrane\cite{renew5}
complexes. This is due to the strong steric repulsion from large
numbers of inclusions, which leads to the uniform dispersion of
inclusions.   Further increase of $\phi_{d}$ leads to the
deformation of lipids which produces the effective attraction
between inclusions.  The deformed conformational entropy can
partially be   released by chaining of inclusions under the
lipid-mediated attraction, shown in Fig. 3e. Such a regularly
modulated inclusion-rich stripe structure has been reported in the
drug-membrane\cite{ne5,ne9} and cholesterol-membrane
complexes\cite{renew5,roger} where the lipids form ribbons between
the aligned cholesterol domains.

\begin{figure}
\setlength{\abovecaptionskip}{3.pt}
\includegraphics[width=9cm]{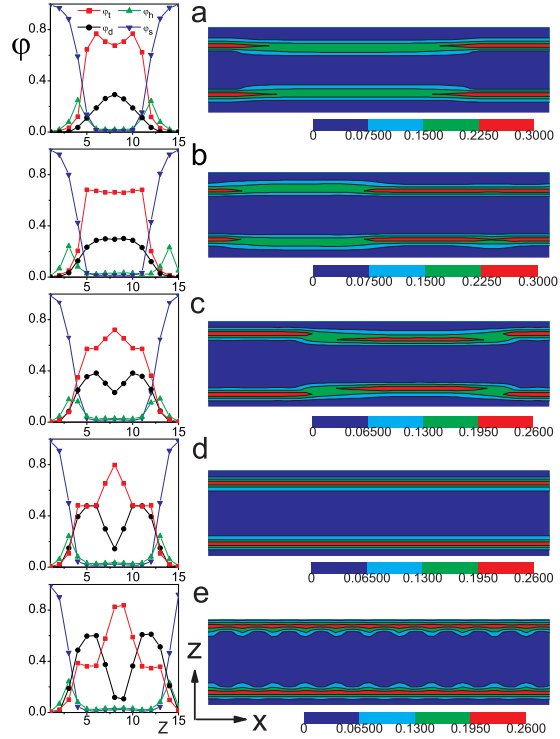}
\caption{(color) Laterally averaged density profiles of
solvent($\varphi_{s}$), lipid tail($\varphi_{t}$),   head
group($\varphi_{h}$), and inclusion ($\varphi_{d}$)   along z axes
(left),  and the x-z cross-sections of density distributions of
lipid heads averaged in the y-direction(right). Parameters are the
same in Fig. 2. }
\end{figure}
Figure 4a shows the translational entropy $S_{d}$ of inclusions.
For low $\phi_{d}$, the large increase of $S_{d}$ with $\phi_{d}$
indicates  that  $S_{d}$ plays an important role at first, which
can account well for the formation of the weak inhomogeneous
distribution of lipids(Fig. 3a) to ensure large translational
entropy of inclusions. For large  $\phi_{d}$,  $S_{d}$ decreases
with the appearance of the chaining structure. Figure 4b shows the
translational entropy $S_{tr}$ of lipids.
  The curve goes up monotonously,
meaning that the addition of inclusions increases the lateral
 membrane fluidity, which is a characteristic of gel-liquid
transition\cite{renew5,ne5}. The reason is that lipid tails are
 stretched along the bilayer normal, which increases  translational degrees of freedom and
 thus enhances the lateral mobility of lipids\cite{ourit1}.
 Figure 4c shows the conformational entropy $S_c$ of lipids. Beginning from a low
conformational entropy in the gel phase of the pure membrane,
lipids get more conformations with the perturbation of added
inclusions where the lipid-enriched rafts are formed.  However,
there is a small decrease in the range $\phi_{d} = 0.24\sim 0.28 $
with the disappearance of lipid rafts, where all lipids are
strongly stretched  with the same length.  Figure 4d shows the
steric repulsion energy $F_{st}$ of inclusions, which increases
with increasing $\phi_d$. Therefore, it is the steric repulsion
between inclusions that suppresses lipid rafts. For  high
 $\phi_d$, the repulsion of inclusions is stronger,
but at the same time, the deformed lipid-mediated attraction
between inclusions becomes also stronger. The only way that the
conformational entropy of deformed lipids is released(Fig. 4c), is
to drive  inclusions to assemble anisotropically along one
direction. As a result, the deformed lipids provide an additional
lateral anisotropic interaction between inclusions, which
stabilizes the parallel chain arrays of inclusions.

Finally, Fig. 5 shows laterally averaged density profiles of
different components across the bilayer  in the left column and
average density distributions of lipid heads in  x-z
cross-sections in the right column. For low $\phi_d$(Fig. 5a), the
inclusions assemble in the bilayer midplane for the membrane
stability, which is displayed by one peak of density profiles
($\varphi_{d}$) of inclusions.  With increasing $\phi_d$, one peak
density profile may be saturated (Fig. 5b), and further increase
of inclusions leads to the occurrence of two peaks of
$\varphi_{d}$ (Fig.5c-e), implying the relocation of inclusions
where the two-layer distribution of inclusions is arranged in
opposing leaflets of a bilayer\cite{renew005}. Previous
experiment\cite{ne5} on drug-membrane complexes has also shown
that the location of drugs in the bilayer depends on drug content.
The two-layer distribution of inclusions   ensures that the ends
of lipid tails can still remain in the membrane midplane favoring
the conformation of lipids. On the other hand, in Fig. 5a-c,   the
irregular density distribution of headgroups originates from the
lipid chain-length mismatch, while  in Fig. 5d, the membrane
surfaces become flat with a matched   length of strongly stretched
lipids where lipid-rich domains disappear. By comparison of the
distances between two peaks of head profiles ($\varphi_{h}$) in
the left side of Fig. 5, we also find that the thickness of
membrane continuously increases with the addition of inclusions.

This work was supported by the National Natural Science Foundation
of China, Nos. 10334020,  20674037, and   10574061.

\end{document}